# A Comparative Study Between a Micromechanical Cantilever Resonator and MEMS-based Passives for Band-pass Filtering Application

Joydeep Basu
Dept. of Electronics and Electrical Communication Engg.
IIT Kharagpur, India
joydeepkgp@gmail.com

Subha Chakraborty
Dept. of Electronics and Electrical Communication Engg.
IIT Kharagpur, India
subharpe@gmail.com

Anirban Bhattacharya
Advanced Technology Development Centre
IIT Kharagpur, India
anirban.bhttchry@gmail.com

Tarun Kanti Bhattacharyya
Dept. of Electronics and Electrical Communication Engg.
IIT Kharagpur, India
tkb@ece.iitkgp.ernet.in

*Abstract—* Over the past few years, significant growth has been observed in using MEMS based passive components in the RF microelectronics domain, especially in transceiver components. This is due to some excellent properties of the MEMS devices like low loss, excellent isolation etc. in the microwave frequency domain where the on-chip passives normally tend to become leakier and degrades the transceiver performance. This paper presents a comparative analysis between MEMS-resonator based and MEMS-passives based band-pass filter configurations for RF applications, along with their design, simulation, fabrication and characterization. The filters were designed to have a center frequency of 455 kHz, meant for use as the intermediate frequency (IF) filter in superheterodyne receivers. The filter structures have been fabricated in PolyMUMPs process, a three-polysilicon layer surface micromachining process.

I. INTRODUCTION

Since the late 1980s, major advances have been seen in wireless communication technologies. In the present wireless communication scenario, numerous standards such as CDMA and GSM exist which provide us voice, data and broadband communication [1]. In order to maintain the quality and reliability of these state of the art technologies, the specifications given to a communication circuit design engineer are getting more and more stringent. The continual adoption of such stringent requirements as demanded by advanced wireless systems such as software defined radios and cognitive radio systems requires development of emerging technologies such as Radio Frequency Microelectromechanical System (RF-MEMS) based devices [2].

The goal of any telecommunication system is the ability to reconfigure the frequency of operation [3]. Traditionally, designers used to avail the on-chip passive components such as pn junction varactors, MOS varactors etc. for such change in the frequency of operation. But as the frequency of operation enters the UHF band, the on-chip passives tend to become leaky and start dissipating more power in the chip. As a result of which, the performance of the transceiver chip declines.

MEMS based passive components play a very important role in this aspect. RF MEMS switches, varactors, inductors and resonators are ideal for reconfigurable systems at GHz range of operation. These components normally possess low insertion loss and very high quality-factor (Q) even up to tens of GHz frequency. Moreover, RF MEMS devices generate very low intermodulation products, a fact which is necessary if the reconfigurable circuit is to be placed before the low-noise amplifier (LNA)/mixer chain. Thus, with the availability of high performance passives over wide frequency range of operation and the immense potential of integration with CMOS based circuitry; RF MEMS based circuits have become an important technology of interest to the wireless design community [4].

Low-loss, high-performance filters are critical components in any transceiver chip [5]. They are widely used as band selection filter, image rejection (IR) filter and intermediate frequency (IF) filter in heterodyne receiver blocks [6]. The performance of such filters is one of the most important governing factors for the performance of the entire receiver chain. Indeed, a great deal of time is spent in designing high-performance filters. One of the main trends of modern radio frequency integrated circuit (IC) design is to develop "single-chip" low cost, low power and miniaturized radio transceivers by integrating all the modules on a single silicon substrate. With the remarkable progress in IC technology, the active circuit blocks have been monolithically integrated. However, most of the commercial designs continue to rely on off-chip components for RF band-pass filtering due to the limited quality factor of the on-chip passives. Off-chip filters like surface acoustic wave (SAW) filters or bulk acoustic wave (BAW) filters provide a better solution as far as the high isolation and low insertion loss are concerned. In particular, the majority of wireless communication transceivers rely mostly upon the high Q mechanical resonators to achieve adequate frequency selection in their RF and IF filtering stages. With recent advancements in MEMS based IC processes targeted at RF designs; MEMS filters were proposed to have capability of signal conditioning applications, where a very high Q value is essential. Also, the manufacturability of CMOS compatible MEMS processes bears the promise of fully integrated RF front-end modules.

The goal of this work is the comparison between two possible MEMS based BPF topologies. The first band-pass filter was realized utilizing MEMS based capacitors and resistors. The other filter was implemented using a micromechanical cantilever resonator having very high Q. Both of these were designed to be used as an IF filter with a centre frequency of 455 kHz [7]. The filters were designed in CoventorWare platform [8] following the PolyMUMPs surface micromachining process [9].

## II. THEORY AND DESIGN

### A. MEMS Passives Based BPF

One of the most common types of frequency-selective filters is a band-pass filter which allows frequencies within a certain range to pass properly from input to the output and blocks frequencies outside that range by imposing adequate attenuation on them. A basic type of BPF can be made using a simple resistance and a capacitance as shown in the fig. 1. The performance of such a frequency-selective filter is described precisely by considering its transfer function $H(s) = V_2(s)/V_1(s)$, which is given by the following expression:

$$H(s) = \frac{sRC}{s^2 R^2 C^2 + 3sRC + 1} \quad (1)$$

where $V_1$ is input voltage and the $V_2$ is output voltage. For a BPF, the range of passband is $f_1 \leq f \leq f_2$ and the rest of the frequencies are in the stopband. The frequencies $f_1$ and $f_2$ are known as the lower and upper 3 dB cutoff frequencies where the gain of the filter falls by 3 dB from its passband value. The centre frequency ($f_0$) of the BPF is given by:

$$f_0 = \frac{1}{2\pi RC} \quad (2)$$

This band-pass filter was implemented using MEMS based variable capacitor and resistor. The micromachined varactor was designed in PolyMUMPs process, which is a three polysilicon layer process [9]. Due to its simplicity of design and low stiffness constant, parallel plate topology was chosen. Since the BPF made up of this varactor is targeted to be integrated with CMOS based circuit components of a transceiver block, reduction in actuation voltage is an important point which needs to be taken care of during design. Lower stiffness constant reduces the pull-in voltage of the varactor and subsequently, the range of actuation voltage is reduced. The three-dimensional view of the varactor is shown in fig. 2.

The varactor structure consists of a rectangular electrode of 0.5μm thickness made up of Poly0 polysilicon layer and a suspended rectangular proof-mass in the Poly1 polysilicon layer having a thickness of 2 μm. There is a 2 μm air gap between these two layers. The proof-mass is a quad-beam structure where four rectangular beams in Poly1 layer are attached to the proof-mass symmetrically on two opposite sides. The entire Poly1 layer comprising proof-mass and the four beams is kept hanging with the help of two sidewall anchors. There are several perforations in the proof-mass. These holes have dual role in this structure. Firstly, the last step of the fabrication process is HF etching which is done to remove all sacrificial oxide layers. These etch holes help in proper release of the structure by allowing HF flow through them at the innermost portion of the proof-mass, where there is very little chance for HF to flow from the side gap of 2 μm. Secondly, these perforations help in reducing squeezed-film air damping during varactor operation. The device dimensions of the varactor are shown in Table 1.

The principle of operation of the varactor is based on electrostatic actuation. When a dc potential is applied to the electrode with the proof-mass grounded, due to electrostatic actuation, the air gap reduces; resulting in an increase in the parallel-plate capacitance. Thus, by changing the applied potential, the structure can be made to operate as a variable capacitor or varactor.

The resistance was designed in PolyMUMPs process and was realized in the Poly0 layer. In order to achieve the desired resistance in minimum required area, the shape was chosen to be serpentine with a uniform thickness of 2 μm [10].

### B. MEMS Resonator Based BPF

An electric-field driven micromechanical cantilever resonator is one of the most fundamental and widely studied structures in MEMS, which can provide a high-Q and narrow band-pass filtering function. This high Q factor is an important figure of merit since it solely determines different properties such as power consumption, loss, noise figure etc. [3]. Although, higher order filters are normally required for getting filter responses with larger bandwidths and flatter pass-bands by coupling separate mechanical resonators via soft mechanical springs [11], we would concentrate on the response of an individual cantilever beam in this paper.

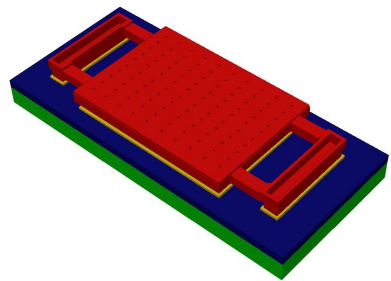

Figure 2. 3D view of the designed varactor

TABLE I   DEVICE DIMENSIONS OF THE VARACTOR

| Component name | Description |
|---|---|
| Actuation electrode | 320μm×220μm×0.5μm |
| Proof-mass | 340μm×240μm×2 μm |
| Beam | 100μm×20 μm×2 μm |
| No. of beams | 4 |
| Holes in the proof-mass | 4μm×4μm×2μm |
| No. of holes | 100 |

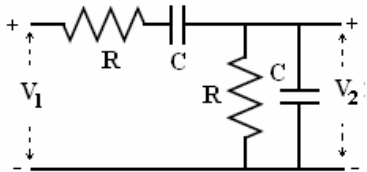

Figure 1. A basic BPF configuration using RC passive

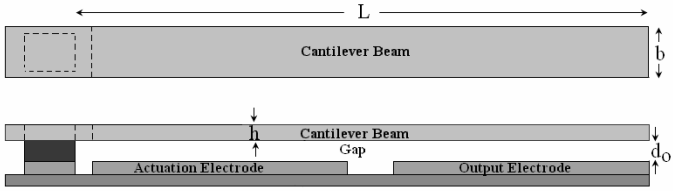

Figure 3. Top view (top) and side view (bottom) of the cantilever resonator.

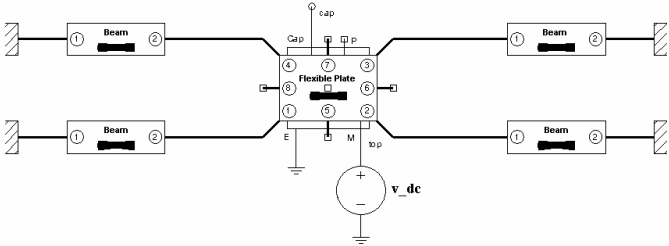

Figure 4. Schematic of the varactor structure in Saber platform.

The schematic diagram of a simple cantilever beam fixed at one end is illustrated in fig. 3. The instantaneous deflection under undamped free vibration of the beam of length $L$, rectangular cross-section with width $b$ and thickness $h$, as shown in the figure, is governed by the Euler-Bernoulli equation [12], [13]

$$EI\frac{\partial^4 y}{\partial x^4} + \rho A \frac{\partial^2 y}{\partial t^2} = 0 \qquad (3)$$

where $y(x,t)$ is the deflection of the beam at a distance $x$ from the clamped end at an instant $t$, $\rho$ and $E$ are density and Young's modulus of the beam material respectively, $A$ is the cross sectional area and $I$ is the area moment of inertia of the beam. Solving this equation, the natural frequency of the beam for $n^{th}$ mode can be determined as [14]

$$\omega_n = \left(\frac{k_n^4}{12}\right)^{\frac{1}{2}}\left(\frac{h}{L^2}\right)\sqrt{\frac{E}{\rho}} \qquad (4)$$

where $k_n$ can be found to be 1.875 [14]. The most dominant damping mechanism that affects the kinematics of the beams is squeezed-film damping and the Q factor associated with it is given by [14]

$$Q_{squeeze} = \frac{2\pi \rho h y_0^3 f_n}{\mu b^2} \qquad (5)$$

where $y_0$ is the initial gap between the cantilever beam and the bottom electrodes and $\mu$ is the dynamic viscosity of air, having a value of $1.81 \times 10^{-5}$ Pa s under normal conditions [14]. From (4) and (5), it is evident that the Q factor is inversely proportional to the square of the overlap area between the cantilever and the electrodes. Since this area is quite small in case of cantilevers, Q factor is normally very high in these structures. Also, rarefaction of air surrounding the cantilevers leads to further increased Q factor.

To properly excite this device, a dc-bias voltage $V_P$ is applied to the beam and an ac excitation $v_i$ is applied to the actuation (or, input) electrode. Thus, a resultant potential difference of ($V_P$-$v_i$) is produced between the beam and input electrode. Hence, a force component is created between this electrode and the beam having the same frequency as that of $v_i$. When the frequency of $v_i$ approaches the beam's fundamental resonance frequency, it begins to vibrate in a direction perpendicular to the substrate, creating a dc-biased time varying capacitor $C_o(t)$ between the beam and the output electrode. As given in (6), $C_o(t)$ has a fixed as well as a sinusoidally varying component.

$$C_o(t) = C_{fix} + C_{var}\sin(\omega t) \qquad (6)$$

For all input frequencies away from the resonant frequency, $C_o \approx C_{fix}$. But, at resonance, we have large deflections of the beam with $C_{var}$ assuming a significant value. Thus, in this condition, an output motional current $i_o$ is produced at the output transducer as given by:

$$i_o = V_P \frac{\partial C_o}{\partial t} = V_P C_{var} \omega \cos(\omega t) \qquad (7)$$

Hence, a frequency filtering function is realized. The current can be directed to a resistor $R_o$, which provides the proper termination impedance for the micromechanical filter. $R_o$ can further feed a transresistance amplifier which amplifies the current to a buffered output voltage $v_o$.

To achieve the desirable resonance frequency of 455 kHz, the required length of the beam was determined from (4) to be 76.70 μm. The cantilever resonator was designed for a Q factor of 50, which led to the width of 10.67 μm from (5).

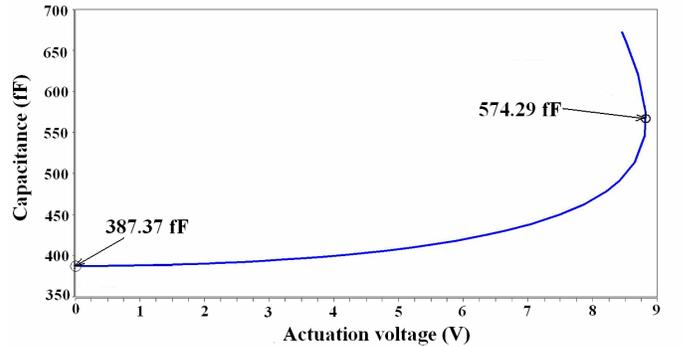

Figure 5. C-V curve of the MEMS based capacitor.

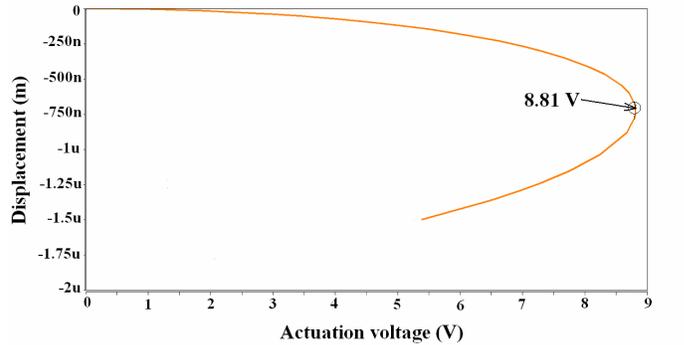

Figure 6. Pull-in curve of the variable capacitor.

Therefore, the width was taken to be 10 μm. The beam was made using the Poly1 layer, while the bottom electrodes were patterned in the Poly0 layer of polysilicon. The thickness of the beam is thus 2 μm. The resulting electrode-to-beam air-gap becomes 2 μm. Electrical connections to the beam and the I/O electrodes were also realized using the Poly0 layer.

III. SIMULATION RESULTS

*A. MEMS Passives Based BPF*

The varactor structure was simulated using Saber Architect in CoventorWare platform using lumped components which can accurately model the behavior of this varactor. The schematic of the capacitor is shown in fig. 4. Firstly, coupled electromechanical simulation has been done to get the C-V curve and the pull-in curve. These two curves are shown in figures 5 and 6 respectively. From those graphs, it is seen that pull-in takes place at 8.81 V and the capacitance varies from 387.3 fF to 574.3 fF when the applied voltage is varied from 0 V to 8.8 V. Therefore, when this varactor will be used in BPF, tunability in the centre frequency can easily be achieved.

In order to determine the performance of the filter, the structure was realized in Saber platform for system level simulation. The schematic of the passive based BPF is given in fig. 7. For a centre frequency of 455 kHz and a capacitance of 387.3 fF, the required value of the resistance is 903 kΩ. In simulation, a resistance of 890 kΩ was used to get the desired centre frequency. The frequency response curve of the filter is shown in fig. 8. From the curve, it can be observed that the 3dB cutoff frequencies of the BPF are 1.48 MHz and 139 kHz respectively. Therefore, with a centre frequency of 455 kHz, the Q of the BPF circuit can be computed as 0.34. The Q can be further improved by using some other BPF topology.

However, there is one advantage of using this passive based BPF. By changing the dc actuation voltage of the varactor, 50% change in the capacitance can be achieved before pull-in [3]. As a result, the centre frequency $f_0$, which is equal to $1/2\pi RC$, can be changed by 50% too. Such a wide tunability in the centre frequency can be achieved by using MEMS varactor based band pass filter.

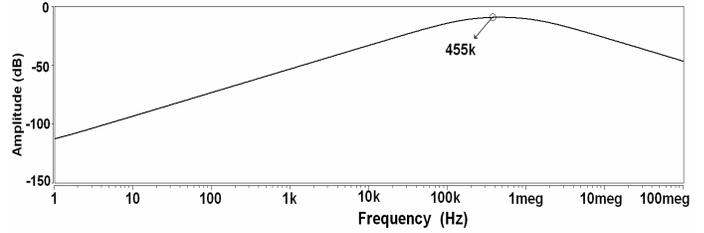

Figure 8. Frequency response of the passive based BPF.

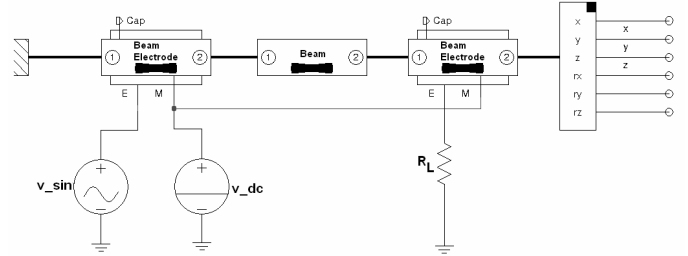

Figure 9. Schematic of the cantilever resonator based BPF.

*B. MEMS Resonator Based BPF*

The 76.7 μm long polysilicon cantilever beam resonator was simulated in CoventorWare and its modal analysis was also performed to determine its mode shapes and the corresponding modal frequencies. The structure was also simulated in Saber Architect platform. The schematic of the structure is shown in fig. 9. The first desirable mode of vibration at 455 kHz is shown in fig. 10. Also, the frequency response characteristic of this beam, found from the harmonic analysis, is shown in fig. 11. From this figure, it can be seen that at 455 kHz, the amplitude of the first mode of vibration is maximum, whereas the amplitudes of the higher order modes are practically negligible. Therefore, at 455 kHz, band-pass characteristics can be achieved by using the fundamental mode of vibration.

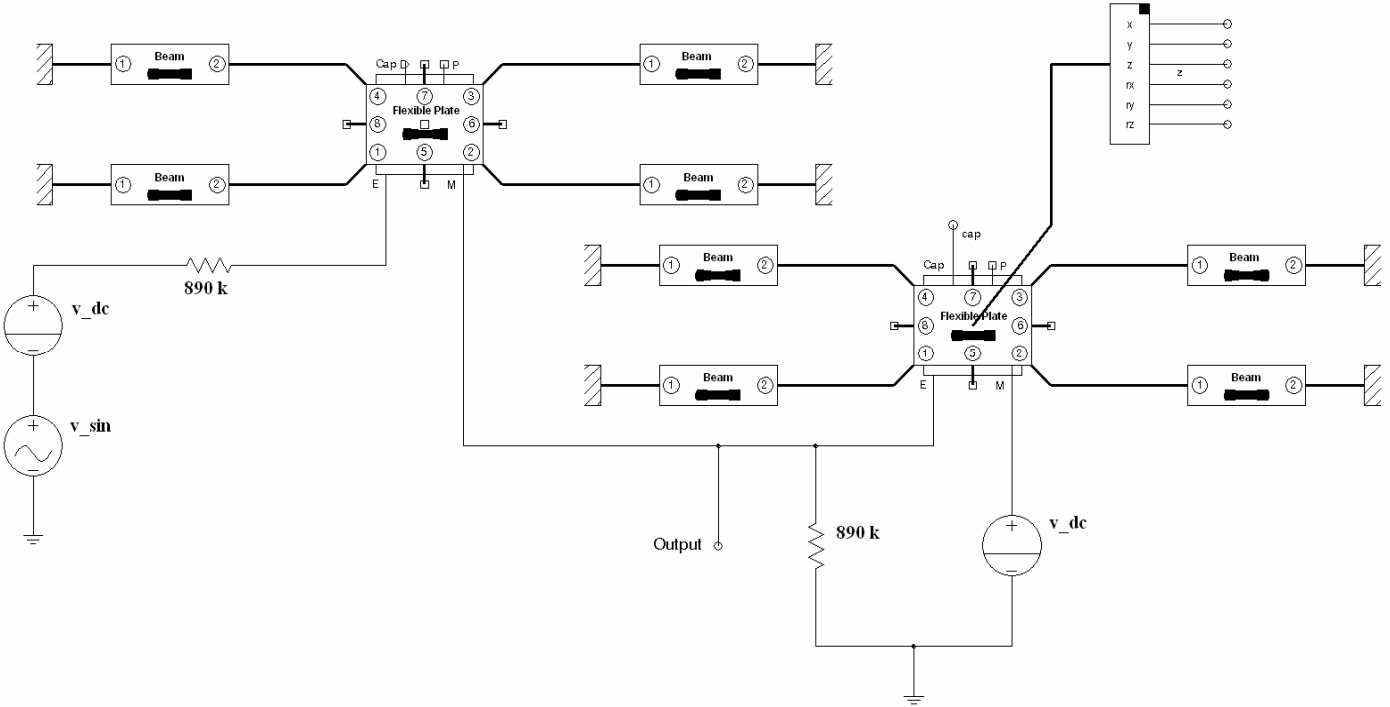

Figure 7. Schematic of the MEMS passive based BPF.

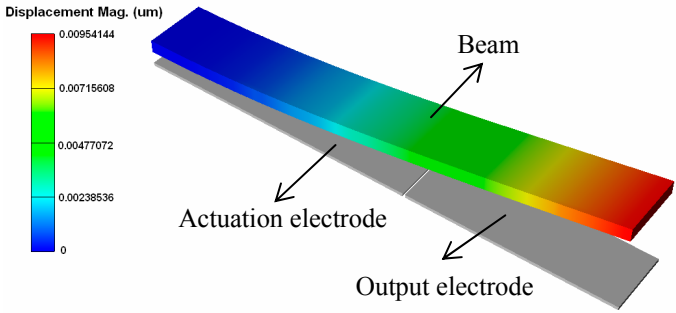

Figure 10.  First mode shape of the cantilever beam at 455 kHz.

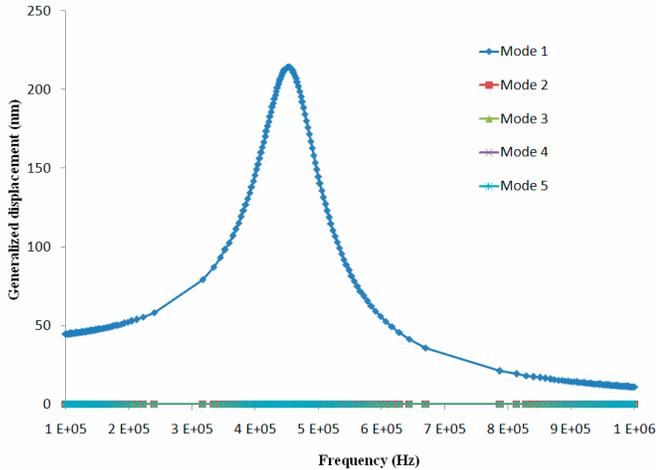

Figure 11.  Frequency response of the cantilever beam.

## IV.  FABRICATION AND TESTING

MEMS passives based BPF structures were fabricated in PolyMUMPs process. The varactor was implemented in Poly0 and Poly1 layers. Hole1 mask was utilized to realize the holes in the proof-mass. The resistance was realized in the Poly0 layer. Pads were used for electrical connection to the bottom electrode and the proof-mass. The 2μm air gap was realized using the first sacrificial oxide layer and then etching it in the last step of fabrication process. The SEM image of the fabricated varactor device is shown in fig. 12.

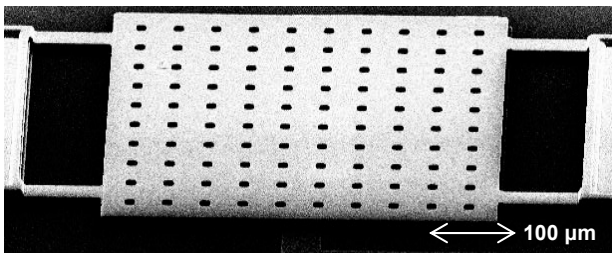

Figure 12.  SEM image of the fabricated varactor structure.

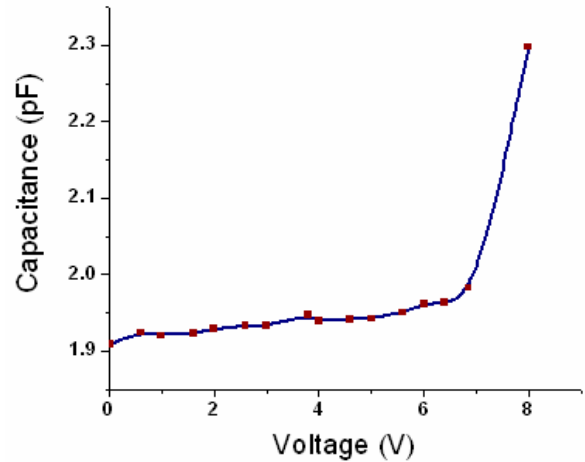

Figure 13.  Measured C-V curve of the varactor device..

The varactor was characterized for C-V profile using LCR meter. The C-V plot is shown in Fig. 13. From that figure, it can be observed that the measured capacitance varies from 1.92 pF to 2.29 pF over the voltage range of 0-8 V. Therefore, it can be seen that in addition to the parallel plate capacitance, an extra parasitic capacitance of about 1.6 pF comes in parallel to it. This parasitic capacitance comes through the nitride layer of the PolyMUMPs process.  As a result, the measured centre frequency of the band-pass filter would be slightly shifted from 455 kHz.

The cantilever structure was fabricated using the same PolyMUMPs process. In order to thoroughly characterize the frequency response of the cantilever based resonator beam, an array of beams of different lengths was fabricated instead of a single beam. The SEM image of the fabricated array is shown in fig. 14. The beams and electrodes were realized using the Poly1 and Poly0 polysilicon layers respectively. Anchor1 was used for fixing one end of the beams to the substrate.

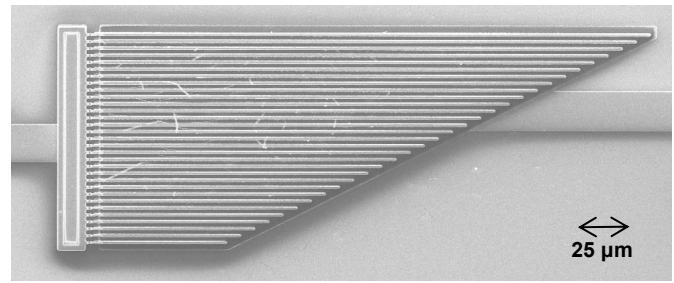

Figure 14.  SEM image of the array of cantilever beams.

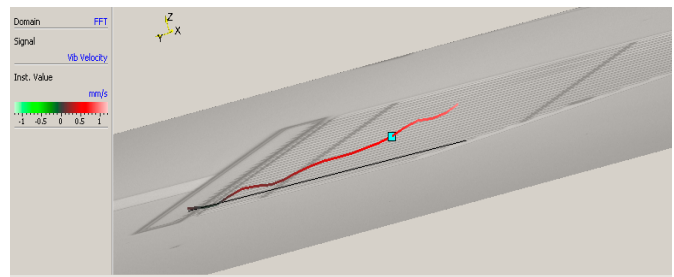

Figure 15.  LDV response of a cantilever beam resoonator of length 80 μm showing the resonant frequency of 441.2 kHz.

The array of cantilevers was electro-mechanically characterized at the wafer level using Laser Doppler Vibrometry (LDV) technique [15]. Wafer prober was used to offer electrical connection to the arrays and sinusoidal voltage signal of frequency sweeping from 0 to 1 MHz was applied at the bottom electrode to excite out of plane modes in the structures. The vibration spectrum of the cantilever beam resonator of 80 µm length is shown in fig. 15. From the figure, it can be seen that the first mode of vibration was at 441.2 kHz. A minor decrease in the fundamental resonance frequency was mainly due to squeezed-film air damping.

## V. CONCLUSION

In this work, we have compared MEMS passive component based and micromechanical cantilever resonator based band-pass filter configurations. From the simulation results and test data, it is evident that the resonator based band-pass filter shows much higher quality factor and frequency selectivity in comparison to its RC based counterpart. Although by selecting narrow BPF topology, the Q of the RC based BPF can be improved, but that would require active elements in the circuit such as a high speed op-amp. For IF operation, such active components can be obtained. But for higher frequency of operation especially in the microwave domain, finding suitable active components with adequate gain and bandwidth becomes really difficult. Also, there is an issue of integrability of the MEMS components with CMOS circuit components. In addition to that, the area requirement of MEMS RC based BPF is much higher than the resonator based BPF. On the other hand, the major advantage of using the RC based filter is its huge tunability. In RC based filter, a maximum of 50% tunability in centre frequency can be achieved whereas a resonator based band-pass filter has much lower frequency selectivity in comparison. Therefore, from VLSI perspective, resonator based filters are preferable while RC based filters would be given more priority when reconfigurability and tunability is the main concern of a filter designer.

## ACKNOWLEDGEMENT

The authors would like to thank National Programme on Micro and Smart Systems (NPMASS), Govt. of India for sponsoring the project. We would like to acknowledge the assistance from Dr. Rudra Pratap, Dr. Navakanta Bhat and their group members for their help in characterizing the devices in their respective laboratories at Indian Institute of Science, Bangalore.